\documentclass[prl, twocolumn, superscriptaddress]{revtex4-1}
\usepackage{amsmath}
\usepackage{amssymb}
\usepackage{graphicx}
\usepackage{xfrac}
\usepackage[crop=pdfcrop]{pstool}
\usepackage{tikz}

\begin{document}
\title{Inverse Ising inference, hyperuniformity and absorbing states in the Manna model}
\author{Thomas Machon}
\email{t.machon@bristol.ac.uk}
\affiliation{H. H. Wills Physics Laboratory, University of Bristol, Bristol BS8 1TL, UK}
\affiliation{Department of Physics and Astronomy, University of Pennsylvania, 209 South 33rd Street, Philadelphia, Pennsylvania 19104, USA}
\begin{abstract}
Using inverse Ising inference we show that the absorbing states of the one-dimensional Manna model can be described by an equilibrium model with an emergent interaction displaying short-ranged repulsion and long-ranged attraction. As the model approaches the critical point the interaction becomes purely repulsive, decaying as $r^{-\alpha}$ and we conjecture the exact value $\alpha=1/2$, suggesting density fluctuations decay as $r^{-3/2}$. We present a simple Gaussian field theory for the long distance behaviour of critical absorbing states and discuss implications for the Manna universality class.
\end{abstract}
\maketitle

For a distribution of particles in a domain of dimension $d$, two extremes of order are uniform randomness and perfect crystalline structure. In the first case, the expected density of particles in regions of size $\ell$, $\langle \rho_\ell \rangle$, may be thought of as the average of $\sim\ell^d$ iid random variables, so that the density fluctuations behave as $\sigma^2(\rho_\ell) \sim \ell^{-d}$. For a perfect crystal the number of particles in a region $R$ is proportional to the number of unit cells in $R$ with variation only in how $\partial R$ ($\sim\ell^{d-1}$) intersects the crystal lattice. It follows for reasonable choices of $R$, such as spheres of radius $\ell$, that the density fluctuations behave as $\sigma^2(\rho_\ell) \sim \ell^{-(d+1)}$. Between these two extremes lie classes of systems which, while disordered, display suppressed density fluctuations where $\sigma^2(\rho_\ell) \sim \ell^{-\lambda}$, $d<\lambda <d+1$; such systems are said to exhibit disordered hyperuniformity~\cite{torquato03,gabrielle02}. In Ref.~\cite{hexner15} a variety of critical absorbing states in non-equilibrium models were shown to have suppressed, hyperuniform, density fluctuations, with similar findings in other studies~\cite{grassberger16,dickman15}.

Because a distribution of absorbing states has no intrinsic dynamics, it is completely characterised by the equilibrium distribution of some unknown (and potentially unphysical) Hamiltonian. The goal of this paper is to determine the form of the equilibrium model corresponding to the critical point of an absorbing phase transition in the Manna universality class. Using inverse Ising inference we numerically estimate a model that reproduces the suppressed density fluctuations found in absorbing states as the density $\rho$ approaches the critical density $\rho_c$. Below $\rho_c$ we find an effective potential exhibiting short-range repulsion and long-range attraction decaying as $r^{-\beta}$, with a minimum located at $x_{min}(\rho_c-\rho)$. As $\rho_c-\rho \to 0$, $x_{min}$ as $(\rho_c-\rho)^{-\gamma}$, where $\gamma \approx 1.4$. 

At $\rho_c$ the interaction is purely repulsive, decaying as $r^{-\alpha}$. Based on our numerical results, we conjecture that $\alpha$ takes the exact value $1/2$. We present a short analysis of the inferred equilibrium model at the critical point. While numerical simulations reproduce the exponent $\lambda \approx 1.425$ found in Ref.~\cite{hexner15}, the model may also be studied analytically, where the long-range behaviour is Gaussian, and one finds $\lambda = 3/2$. Taken together, the results suggest that the discrepancy may be a finite-size effect, with $3/2$ the true value. This supports the hypothesis~\cite{ledoussal15} that the Manna universality class (believed to be conserved directed-percolation~\cite{basu12}) is related to theories of interface depinning in quenched disorder, for which one would expect $\lambda = 3/2$~\cite{grassberger16}.

\begin{figure}
\includegraphics{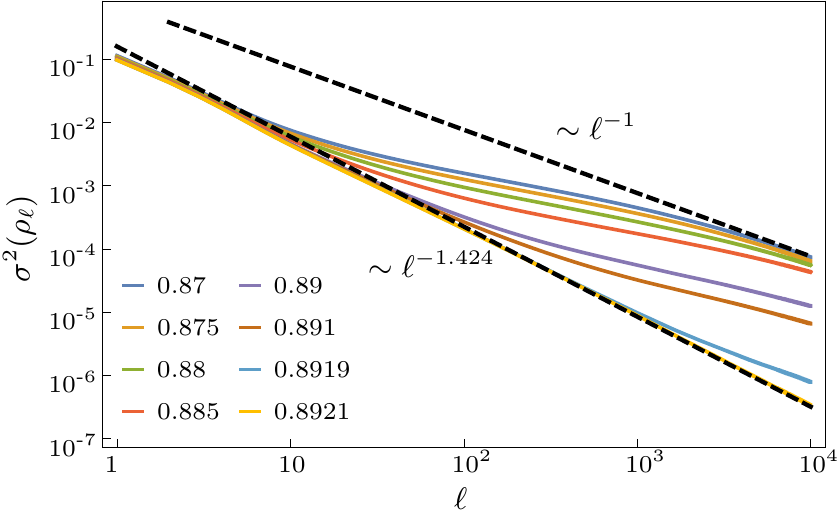}
\caption{Density fluctuations, $\sigma^2(\rho_\ell)$ as a function of density $\rho$ and window size $\ell$ for the one-dimensional Manna model with carrying capacity one. Below $\rho_c$, $\sigma^2(\rho_\ell)~\sim \ell^{-\lambda}$ for small $\ell$ and $\sim \ell^{-1}$ for large $\ell$. As $\rho \to \rho_c$, the crossover no longer occurs. Data from absorbing states on lattices of size $L=10^5$. Sample size $N=100$ except for $\rho=0.8919$ and $\rho=0.8921$ where $N=40$. The data for $\rho=0.8921$ give $\lambda = 1.424 \pm 0.016$.}
\label{fig:0}
\end{figure}

We begin with the one-dimensional Manna model with carrying capacity one~\cite{manna91}. The model is defined on a one-dimensional periodic lattice of size $L$ with each lattice site having occupancy $n_i \in \mathbb{N}_0$. Sites with $n_i$ greater than the carrying capacity are deemed active. Active sites update by distributing all particles amongst their nearest neighbours. In this study these updates are done sequentially with active sites chosen at random, though the model may be defined with parallel updates with no consequence for the critical behaviour~\cite{lee14}. The model thus defined has two parameters, the density $\rho$ and lattice size, $L$. As $L \to \infty$ the model exhibits an absorbing phase transition~\cite{henkelbook}. For $\rho > \rho_c \approx 0.892$~\cite{lee14,basu12,lubeck03,lubeck04} the system is in the active state, and the density of active sites approaches a non-zero constant. For $\rho<\rho_c$ the system eventually reaches an absorbing state, with zero active sites. The Manna model with carrying capacity one was not studied in Ref.~\cite{hexner15}, so we first establish that the absorbing states exhibit disordered hyperuniformity as $\rho \to \rho_c$. The results are shown in Fig.~\ref{fig:0}, and the data for $\rho=0.8921$ give $\lambda = 1.424 \pm 0.016$. This is consistent with the value $\lambda = 1.425 \pm 0.025$ found for other non-equilibrium one-dimensional systems~\cite{hexner15}.

\begin{figure}[t]
\begin{center}
\includegraphics{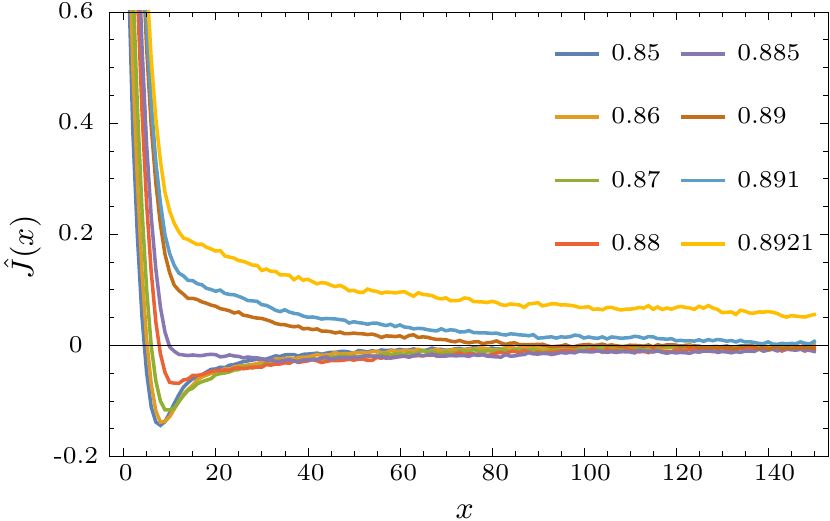}
\end{center}
\caption{Inferred interaction potentials $\hat J(x)$ for the one-dimensional Manna model with varying values of $\rho \in [0.85, 0.8921]$, below $\rho_c \approx 0.8921$, displaying short-range repulsion and long-range attraction. As $\rho \to \rho_c$ the minimum of the potential tends to $\infty$ and the interaction becomes purely repulsive at $\rho_c$ (see Fig.~\ref{fig:crit}). Data shown are from inferred potentials with maximum range $r=1000$. Plots obtained using 100 samples of size $10^5$ for each value of $\rho$ with a range $r=250$.}
\label{fig:1}
\end{figure}

Because the set of absorbing states has no dynamics associated to it, it is completely characterised by the properties of a corresponding equilibrium system. For a given distribution of initial conditions $I$, lattice size $L$ and density $\rho$ the model defines a probability distribution
\begin{equation}
P_{L,\rho,I}\big ( \{ n_i \} \big )
\end{equation}
over valid absorbing states. Given this distribution, one obtains an effective Hamiltonian $H_{L,\rho,I}\big ( \{ n_i \} \big )= -\log P_{L,\rho,I}\big ( \{ n_i \} \big )$. Our goal is to study the limit of this Hamiltonian as $L \to \infty$ near the critical density $\rho_c$. The relaxation timescale diverges as $\rho$ approaches $\rho_c$ and so we assume the dependence of $H_{L,\rho,I}$ on the initial conditions $I$ to be negligible in this regime, and to vanish as $\rho \to \rho_c$. We will denote the limiting Hamiltonian by $H_\rho$ or $H$ if no confusion may arise. The absorbing states of the model have occupancy $n_i \in \{0,1\}$, and so the distribution $P_{L,\rho,I}\big ( \{ n_i \} \big )$ is over configurations of a one-dimensional lattice gas.

While in general $H_{L,\rho,I}\big ( \{ n_i \} \big )$ may contain arbitrary $n$-body interactions, we are only concerned with reproducing the density fluctuations which are determined entirely by the connected correlation function, $g(r)$, as
\begin{equation}
\sigma^2({\rho_\ell}) = \frac{g(0)}{\ell} + \frac{2}{\ell} \sum_{k=1}^{\ell-1}\left (1-\frac{k}{\ell} \right) g(k).
\end{equation}
It is hence sufficient for our equilibrium model to reproduce $g(r)$. The maximum entropy model reproducing an arbitrary two-point correlation function is an Ising-type model, containing only one- and two-body interactions~\cite{jaynes57,schneidman06}. The symmetries of the Manna model impose translational and reflection invariance on $P_{L,\rho,I}\big ( \{ n_i \} \big )$ and so we assume that the equilibrium model has the form
\begin{equation}
H=\frac{1}{2} \sum_{i,j} J(|i-j|) n_i n_j - \mu \sum_i n_i,
\label{eq:eff}
\end{equation}
where $J(0)=0$. Note that such a model may well explore configurations that are not reachable as absorbing states in the Manna model this is acceptable as we are only concerned with reproducing the correlation function, rather than the distribution in full, and ultimately in the long-distance behaviour of model. 

\begin{figure}
    \centering
    \begin{tikzpicture}
        \node[anchor=south west,inner sep=0] (image) at (0,0) {\includegraphics{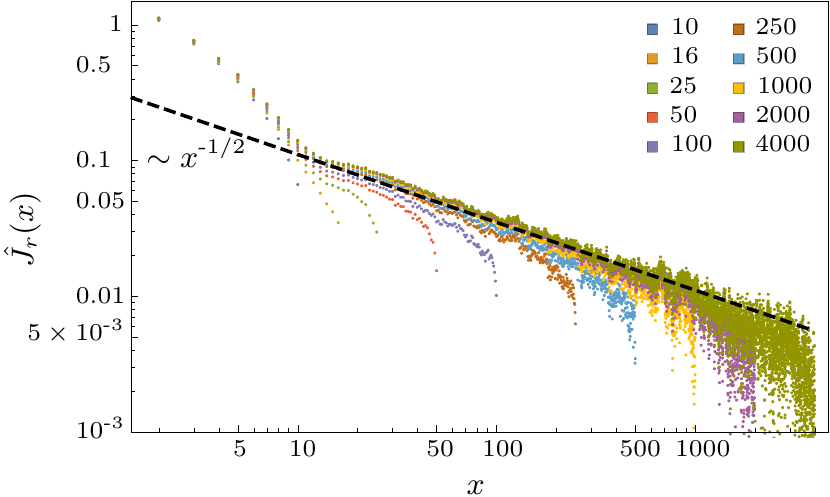}};
            \node[anchor=south west,inner sep=0] (image) at (1.5,0.75) {\includegraphics{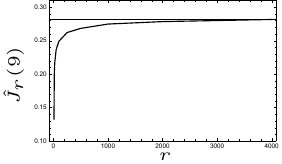}};
    \end{tikzpicture}
\caption{Log-log plot of inferred interaction potential $\hat J_r(x)$ for maximum range $r$ between $10$ and $4000$ close to the critical point, with $\rho =0.8921$, where the interaction is purely repulsive. The black line shows a plot of $0.35 x^{-1/2}$. Inset:  $\hat{J}_r(9)$ as a function of maximum range $r$, illustrating the convergence of the inference process.}
\label{fig:crit}
\end{figure}

We treat this problem numerically, where it falls under the domain of inverse Ising inference~\cite{nguyen17,aurell12}. We take numerically generated configurations and infer the effective interaction model \eqref{eq:eff}. We impose a maximum range, $r$, on the coupling coefficients which we infer, so that $J_r(x)=0$ for $x>r$. In the work presented here $r \ll L$, the lattice size, so we assume any finite-size effects coming from the lattice are negligible compared to the noise in the inference and to finite-size effects from $r$. To estimate the chemical potential $\mu_r$ and the vector of interactions ${\bf J}_r$ we found pseudo-likelihood maximisation (ordered logistic regression) to be effective~\cite{aurell12}. Given a sample $S$ with configuration $\{n_{i,S}\}$ of carrying capacity $C$, and a site $j$, we form the conditional probability
\begin{equation}
P_{j,S} = P\left (n_{j,S} | \{n_{i,S}\} _{\setminus {j,S}} \right) = \frac{z^{n_{j,S}}}{\sum_{k=0}^C z^k},
\end{equation}
where $z=\exp(-H_{j,S})$ and
\begin{equation}
H_{j,S} = \mu_r + \sum_{i=-r}^r J_r(|i-j|)n_{i,S}
\end{equation}
is the contribution to the energy from site $j$ in sample $S$. The estimates $\left ( \hat{\mu}_r,\hat{\bf J}_{r} \right)$ are found by maximising the pseudo-log-likelihood
\begin{equation}
\mathcal{L} = \sum_{S,i} \log P_{S,i},
\label{eq:ll}
\end{equation}
where the sum is over all samples, $S$, and sites, $i$. One can show that if the data are generated by an equilibrium model with two-body interactions, then this procedure gives a consistent estimator of the true interactions~\cite{aurell12}. We used BFGS pseudo-Newtonian optimisation to maximise \eqref{eq:ll}. To ensure stability of the algorithm for large $r$, inferred couplings for $\tilde r< r$ were used as initial conditions to estimate $\left ( \hat \mu_r,\hat {\bf J}_{r} \right)$. For this process to be well defined, the estimates of $\hat {\bf J}_{r} $ should converge as $r$ gets large (note that $\hat \mu_r$ need not converge for reasons discussed below) and this is indeed the case (see Fig.~\ref{fig:crit}).

Results for the inferred interaction potentials for $\rho$ close to to $\rho_c$ are shown in Fig.~\ref{fig:1} (similar results were obtained for other carrying capacities). For $\rho<\rho_c$, the interaction potential displays short range repulsion and long-range attraction with the minimum of the interaction located at some value $x_{min}(\rho)$. As $\rho \to \rho_c$, $x_{min} \to \infty$ so that the interaction is purely repulsive at the critical point, consistent with the negative correlation functions that are typical of hyperuniform systems~\cite{torquato03}. We note also that other non-equilbrium systems with local dynamics can lead to effective long-range equilibrium interactions~\cite{evans98}.

\begin{figure}
\begin{center}
\includegraphics{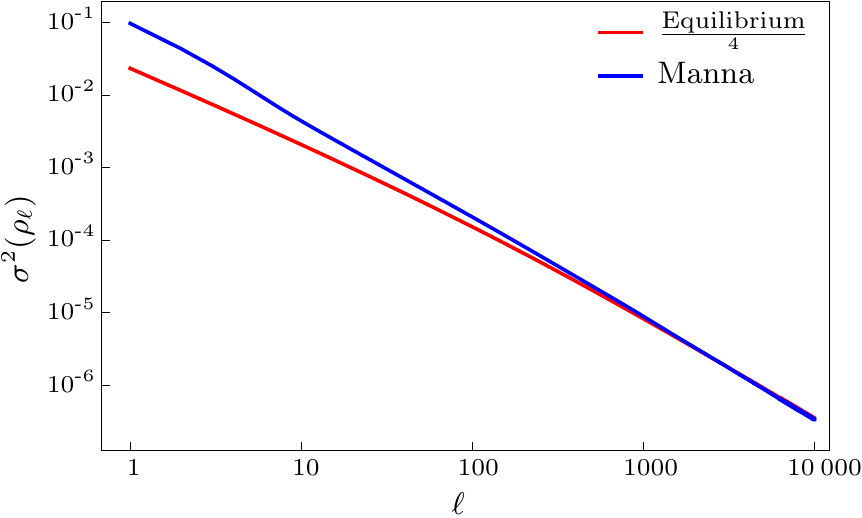}
\end{center}
\caption{Comparison of density fluctuations in the 1D Manna Model for $\rho=0.8921$, and the equilibrium system \eqref{eq:ttt} with $L=10^6$, $\rho = 0.892$ and $J_0 = 0.35$, approximately the value of $J_0$ found in the inference process. While the equilibrium model reproduces the scaling in the density fluctuations, we find that the magnitude of the asymptotic amplitudes do not match, and as plotted, the density fluctuations for the equilibrium model are reduced by a factor of $4$ to allow for better comparison. This discrepancy may be because \eqref{eq:ham} does not include the short-range deviation from power-law behaviour in the inferred potential.}
\label{fig:comp}
\end{figure}

By construction, the inferred potentials reproduce the observed correlations in the absorbing states (see Fig.~\ref{fig:comp}), and it is natural to ask which properties of the Manna model's absorbing states emerge from the statistical mechanics of the inferred equilibrium system, and which are encoded into the parameters of the potentials. A possible example of the latter is given my $x_{min}(\rho)$. As $\rho \to \rho_c$ we expect a scaling
\begin{equation}
x_{min}(\rho) \sim (\rho_c-\rho)^{-a}.
\end{equation}
One may expect $x_{min} \propto \xi_\times$, where $\xi_\times$ is the crossover distance in the density fluctuations so that $a$ should be equal to $\nu_\perp =1.347\pm0.091$, which controls the divergence of the active site-active site correlation length the in the active state~\cite{hexner15,lubeck03}. Computing $x_{min}$ for the inferred potentials, we estimate $a = 1.4 \pm 0.1$, consistent with the value of $\nu_\perp$. We can also study the attractive tail for $\rho<\rho_c$, however the noise inherent in the inference process precludes a precise analysis. The data suggest that the tail decays as $-B x^{-C}$, where both $B$ and $C$ are functions of $\rho_c-\rho$. We estimate $B \sim (\rho_c-\rho)^c$, $c=1.7 \pm 0.3$ and $C \sim (\rho_c-\rho)^b$, $b = 0.53 \pm 0.10 $, with proportionality constant such that $C = 1.46 \pm 0.04$ for $\rho = 0.85$.

We now turn to the purely repulsive behaviour at $\rho_c$, where we expect the model to describe the universal behaviour of the non-equilibrium critical point. We note here that the inference process does not reproduce the distribution of absorbing states of the Manna model, just their correlations, so that only at large lengthscales do we expect their behaviours to match precisely. The required correlations cannot occur with short-range interactions~\cite{hexner15}, and the interaction potential must satisfy (in one dimension)
\begin{equation}
\lim_{s \to \infty} \sum_{i=1}^s J(i) = \infty.
\label{eq:lr}
\end{equation}
If this does not hold then the system is additive and density fluctuations will typically scale asymptotically as $\ell^{-1}$. We discuss some of the subtleties associated with this long-range interaction below. 

The inference procedure is noisy~\cite{note_reg} at long ranges, making direct estimation of the long-range behaviour difficult. Instead, we analyse the behaviour of the chemical potential. In order for the equilibrium model to be well-defined in the thermodynamic limit, the energy difference upon adding a particle to a configuration with the equilibrium density should be finite,
\begin{equation}
\left | H(n_j=1 ,\{n_i\} _{\setminus j} )-H(n_j=0 , \{n_i\} _{\setminus j} ) \right | < \infty.
\label{eq:cond}
\end{equation}
If we assume an infinite chemical potential of the form $\mu = \nu \sum_{i}J(i) $, this condition becomes
\begin{equation}
|\Delta H | = \left | \sum_{i=1}^\infty J(i)\left (n_{j+i}+n_{j-i} -\nu \right ) \right | < \infty.
\label{eq:thingy}
\end{equation}
For an interaction satisfying \eqref{eq:lr}, $\Delta H = + \infty$ if $2\rho> \nu$ and $\Delta H = - \infty$ if $2 \rho < \nu$. It follows that configurations with support in the equilibrium distribution of $H$ must have $2\rho = \nu$, which is a necessary (but not sufficient) condition for \eqref{eq:thingy} to hold; the infinite chemical potential balances the strength of the long-ranged repulsive interaction giving a non-zero equilibrium density.

\begin{figure}
\begin{center}
\includegraphics{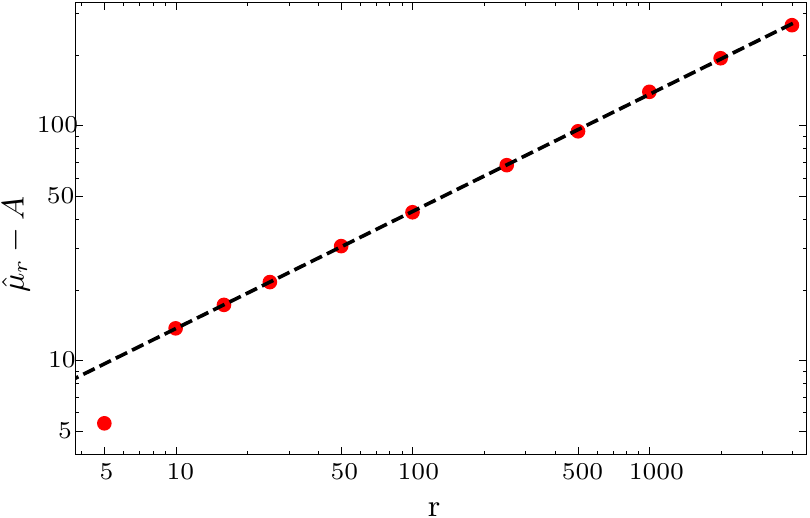}
\end{center}
\caption{A fit of $\hat{\mu}$ to the form $A+B r^C$. Note that $\hat \mu_r - A$ is shown on the vertical axis. We find $C = 0.499 \pm 0.016$.}
\label{fig:3}
\end{figure}

This allows us to infer the long-range nature of $J(x)$. Consider the estimated values $\left ( \hat \mu_r,\hat {\bf J}_{r} \right)$. If we assume for large sample number, that as $r \to \infty$ the estimates of ${\bf J}_r$ converge to $ {\bf J} $ and that $J(x) \sim x^{-\alpha}$ then the asymptotic relation
\begin{equation}
\hat \mu_r \sim r^{1-\alpha}
\end{equation}
should hold as $r$ gets large. Since the inferred interaction has some short-range behaviour that is not a power-law, we add an additional constant and fit the inferred chemical potentials to the form ${\hat \mu}(r) = \mu_0+Br^C$, shown in Fig.~\ref{fig:3}, along with data for ${\hat \mu}(r)$. We find $C = 0.499 \pm 0.016$. Given the strength of this relationship, \emph{we conjecture that the true value for the exponent $\alpha$ is exactly $1/2$}. This analysis permits the consistency check:
\begin{equation}
\hat \rho := \frac{1}{2} \lim_{r \to \infty}\sfrac{ {\hat \mu_r}}{ \sum_{i=1}^r \hat J_r(i)} = \rho.
\end{equation}
$\hat \rho$ may be estimated and compared with the known value of $\rho$. We find for $r=4,000$ and $\rho = 0.8921$, $\hat \rho = 0.90$.

Given this conjecture, we are motivated to explore the equilibrium model so defined. As discussed below, the groundstate of the model may be effectively computed, however, to ensure that the groundstate energy is bounded from below, we add a constant term to the Hamiltonian bringing it to the form
\begin{equation}
H = \frac{1}{2}\sum_{ij}(n_i - \rho) J_{ij} (n_j - \rho) =  \frac{1}{2}\sum_{ij} \sigma_i J_{ij} \sigma_j,
\label{eq:ham}
\end{equation}
where $J_{ij} = J(|i-j|) \sim J_0 |i-j|^{-1/2}$. As established above, the long-range nature of the interactions fixes the density of the system to be equal to $\rho$, and the theory is defined entirely in terms of the fluctuations $\sigma_i = n_i - \rho$. It is important to note that \eqref{eq:ham} is {\em not} the same as the inferred model, shown in Fig.~\ref{fig:crit}, where there is a short-range deviation from power-law behaviour which we do not include in \eqref{eq:ham}.

As a system with repulsive convex interactions the groundstate of the system can be calculated exactly for a given density $\rho$, using the algorithm given by Hubbard~\cite{hubbard78}. As an example, for $\rho=1/n$, the groundstate contains alternating single occupied and $(n-1)$ unoccupied sites with a per site energy of
\begin{equation}
\frac{J_0}{n^2} \left ( \sqrt{n}-1 \right ) \zeta \left ( \frac{1}{2} \right ).
\end{equation} 
Despite the non-integrability of $1/\sqrt{r}$ interaction \eqref{eq:lr} the groundstate of the system is extensive (a similar observation has been made in the case of frustrated systems~\cite{giuliani06}). Ultimately in this case, extensivity of the groundstate is achieved via the infinite chemical potential, which should be set to scale as $L^{1/2}$ for a finite system. The devil's staircase phenomena found for the antiferromagnetic Ising model~\cite{bak82} does not occur here, the strength of the non-integrable interactions locks the density of the system to $\rho$.

We first investigate \eqref{eq:ham} numerically. We simulate the model in the canonical ensemble using standard Monte Carlo methods, the long-range interactions set the density equal to $\rho$, so that only particle-hole swaps are allowed. On a periodic lattice of size $L$, provided the density of the system is equal to the parameter $\rho$, the divergent part of the interactions can be summed out resulting in an effective Hamiltonian
\begin{equation}
H = \frac{J_0}{2L^{1/2}}\sum_{ij}(n_i - \rho) \Phi \left (|i-j|/L \right ) (n_j - \rho),
\label{eq:ttt}
\end{equation}
where for $x \in (0,1)$, using Hurwitz's formula,
\begin{equation}
 \Phi(x) = -2 \zeta \left (\frac{1}{2} \right) +2 \sum_{s=1}^\infty \frac{\cos (2 \pi s x)}{s^{1/2}}
\end{equation}
and we define $\Phi(0)=\Phi(1)=0$. $H$ diverges if the density does not match $\rho$. Fig.~\ref{fig:comp} shows a comparison between density fluctuations in the equilibrium system \eqref{eq:ttt} and those of the Manna model close to the critical point ($\rho=0.8921$). For parameter values $J_0=0.3$~\cite{note_size}, $\rho=0.892$ we find $\lambda = 1.431\pm0.016$. This is consistent with the numerical results from the Manna model, with a comparison of the density fluctuations shown in Fig.~\ref{fig:comp}.

\eqref{eq:ham} may also be investigated analytically. Our interest is in the long-distance behaviour of the model. Consider the variable $\phi_i$, equal to $\sigma$ coarse-grained over regions of size $\ell$. We may write an effective theory
\begin{equation}
H_\phi =   \frac{1}{2}\sum_{ij} \phi_i J^\prime_{ij} \phi_j+ \sum_i P(\phi_i),
\end{equation}
where $\phi$ is now real-valued and $P$ is a polynomial in $\phi$. The form of the original interaction implies that $J^\prime \sim r^{-1/2}$ at long distances ($\gg \ell$). Taking the Fourier transform, the interaction can be written to leading order as $J^\prime(k) \approx J_0 |k|^{-1/2}$. The strength of the interaction means that, under rescaling, the polynomial terms as well as higher order terms in the expansion of $J(k)$ are irrelevant, and so we are led to the scale invariant Gaussian theory
\begin{equation}
H = \int_{BZ} \frac{1}{|k|^{1/2}} |\phi(k)|^2 dk,
\label{eq:ld}
\end{equation}
describing the long distance behaviour of critical absorbing states in the Manna universality class, for which elementary scaling arguments give $\lambda = 3/2$. We therefore conclude $\lambda=3/2$ in \eqref{eq:ham}. 

\begin{figure}[t]
\begin{center}
\includegraphics{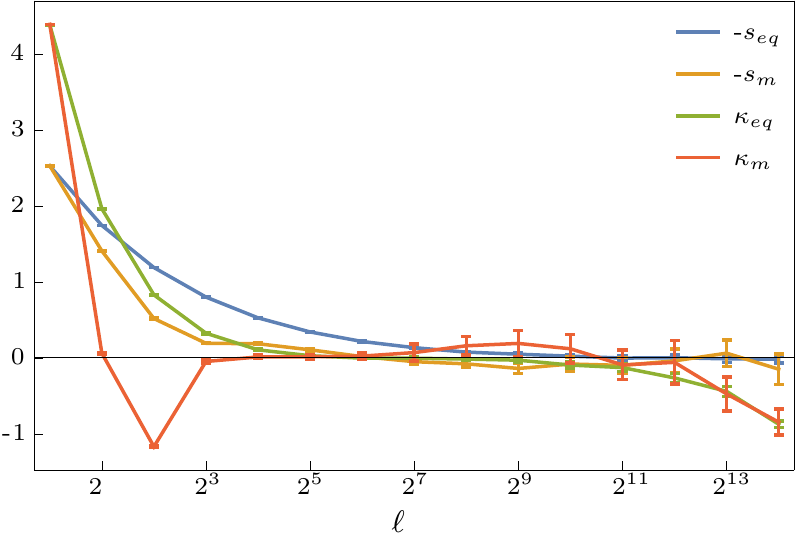}
\end{center}
\caption{Skewness and excess kurtosis of $\rho_\ell$ for the equilibrium model \eqref{eq:ham} and the Manna model (denoted by $eq$ and $m$ respectively). For $\ell \lesssim 2^{12} \sim L/20$, the moments decay to zero suggesting Gaussian behaviour at large distances, supported by more comprehensive statistical tests (Shapiro-Wilks and Anderson-Darling, not shown). Beyond this lengthscale, we see finite size effects in the fourth moment. Uncertainty is $2 \sigma$. Manna data $L=10^5$, $N=40$, $\rho = 0.8921$, equilibrium data, $L=10^5$, $N=600$, $\rho = 0.8921$, $J_0 = 0.2$.}
\label{fig:6}
\end{figure}

Now we must reconcile the numerical value of $\lambda \approx 1.43$ with the analytical value of $\lambda=3/2$ for the equilibrium model. From \eqref{eq:ld}, we expect $\rho_\ell$ to be scale invariant on large lengthscales, so that its standardised moments are constant, and Gaussian. Fig.~\ref{fig:6} shows skewness and excess kurtosis of $\rho_\ell$ for both the equilibrium model as well as Manna model. For the system sizes studied, when $\ell \lesssim L/20$ this is indeed the case, and the behaviour of $\rho_\ell$ is Gaussian for both the equilibrium and Manna models, suggesting \eqref{eq:ld} is accurate (this is confirmed by more robust statistical tests). This no longer holds as $\ell \gtrsim L/20$, where finite size effects become relevant. This can also be seen in the density fluctuations, which show clear finite size effects for $\ell \gtrsim L/10$. While density fluctuations are suppressed on lengthscales $\sim L/2$ for any periodic system, the long-range interactions of \eqref{eq:ham} increase the magnitude of this effect. We note that at fixed system size, simulations of $\eqref{eq:ham}$ at other parameter values given values of $\lambda$ depending on $J_0$ and $\rho$, tending to $3/2$ as $J_0$ gets large and $\rho \to 1/2$, so that the strength of the finite-size effects depend on both $J_0$ and $\rho$. 

Returning to the Manna model, the -$1/2$ interaction obtained by the inverse Ising inference process is well-supported, evident even for small ranges of inferred interactions. If we accept that \eqref{eq:ham} reproduces the long-distance behaviour of the Manna model, as it must, then we have $\lambda = 3/2$ for the Manna model. The conclusion is then that the value $\lambda \approx 1.424$, and other similar numerical values, are due to finite-size effects, which can be large in sandpile models~\cite{grassberger16}. It has been proposed~\cite{ledoussal15} that conserved directed-percolation~\cite{dickman98,vespignani98} (believed to describe the Manna model~\cite{basu12}) is related to the quenched Edwards Wilkinson model~\cite{paczuski96}. If this is the case, then one would expect $\lambda = 3/2$, as has been found for the Oslo model~\cite{grassberger16}. From this perspective then, one may interpret our results as evidence for the hypothesis of Ref.~\cite{ledoussal15}. Finally, we note that an extension to higher dimensions, as well as a derivation of the -$1/2$ exponent would constitute interesting future directions. 

\begin{acknowledgements}
It is a pleasure to acknowledge many useful conversations with S.A. Ridout, who also provided assistance with some computational aspects of this work, as well as helpful discussions with M.O. Lavrentovich and C.W. Lynn. This work was partially supported by the NSF through grant DMR-9732963.
\end{acknowledgements}


\begin{thebibliography}{99}
\bibitem{torquato03} S. Torquato and F. H. Stillinger, Phys. Rev. E {\bf 68}, 041113 (2003).
\bibitem{gabrielle02} A. Gabrielli, M. Joyce, and F. S. Labini, Phys. Rev. D {\bf 65}, 083523 (2002).
\bibitem{hexner15} D. Hexner and D. Levine, Phys. Rev. Lett {\bf 114}, 110602 (2015).
\bibitem{grassberger16} P. Grassberger, D. Dhar, and P.K. Mohanty, Phys. Rev. E {\bf 94}, 042314 (2016).
\bibitem{dickman15} R. Dickman and S.D. da Cunha, Phys. Rev E {\bf 92}, 020104(R) (2015).
\bibitem{ledoussal15} P. Le Doussal and K.J. Wiese, Phys. Rev. Lett. {\bf 114}, 110601 (2015).
\bibitem{basu12} M. Basu, U. Basu, S. Bondyopadhyay, P. K. Mohanty, and H. Hinrichsen, Phys. Rev. Lett. {\bf 109}, 015702 (2012).
\bibitem{manna91} S.S. Manna,  J. Phys. A: Math. Gen. {\bf 24}, L363 (1991).
\bibitem{lee14} S.B. Lee, Phys. Rev. E {\bf 89}, 060101 (2014).
\bibitem{henkelbook} M. Henkel, H. Hinrichsen, and S. Lübeck, {\it Non-Equilibrium Phase Transitions - Volume 1: Absorbing Phase Transitions} (Springer, New York, 2008).
\bibitem{lubeck03} S. L\"{u}beck and P. C. Heger, Phys. Rev. E {\bf 68}, 056102 (2003).
\bibitem{lubeck04} S. L\"{u}beck, Int. J. Mod. Phys. B {\bf 18}, 3977 (2004). 
\bibitem{jaynes57} E.T. Jaynes, Phys. Rev. {\bf 106}, 62 (1957).
\bibitem{schneidman06} E. Schneidman, M.J. Berry II, R. Segev and W. Bialek, Nature {\bf 440}, 1007 (2006).
\bibitem{nguyen17} H.C. Nguyen, R. Zecchina, J. Berg, Adv. Phys. {\bf 66}, 197 (2017).
\bibitem{aurell12} E. Aurell, M. Ekeberg, Phys. Rev. Lett. {\bf 108}, 090201 (2012).
\bibitem{evans98} M.R. Evans, Y. Kafri, H.M. Koduvely, and D. Mukamel, Phys. Rev. E {\bf 58}, 2764 (1998).
\bibitem{note_reg} It is potentially viable to use some form of regularisation here, however it was deemed unnecessary for our purposes.
\bibitem{hubbard78} J. Hubbard, Phys. Rev. B {\bf 17}, 494 (1978).
\bibitem{giuliani06} A. Giuliani, J.L. Lebowitz, E. H. Lieb, Phys. Rev. B {\bf 74}, 064420 (2006)
\bibitem{bak82} P. Bak and R. Bruinsma, Phys. Rev. Lett. {\bf 49}, 249 (1982).
\bibitem{note_size} $J_0=0.3$ is slightly lower than the value $J_0 \approx 0.35$ found by the inference process. 
\bibitem{dickman98} R. Dickman, A. Vespignani, and S. Zapperi, Phys. Rev. E
{\bf 57}, 5095 (1998).
\bibitem{vespignani98} A. Vespignani, R. Dickman, M. A. Mu\~{n}oz, and S. Zapperi,
Phys. Rev. Lett. {\bf 81}, 5676 (1998).
\bibitem{paczuski96} M. Paczuski and S. Boettcher, Phys. Rev. Lett. {\bf 77}, 111 (1996).


\end{thebibliography}
\end{document}